\documentclass[12pt]{elsart}
\usepackage{epsfig}
\usepackage{graphics}
\usepackage{longtable}
\begin{document}
\begin{frontmatter}
\title{Measurement of $\phi$(1020) meson leptonic width
with CMD-2 detector at VEPP-2M Collider}
\author[BINP]{R.R.Akhmetshin},
\author[BINP,NSU]{V.M.Aulchenko},
\author[BINP]{V.Sh.Banzarov},
\author[BINP]{L.M.Barkov},
\author[BINP]{S.E.Baru},
\author[BINP]{N.S.Bashtovoy},
\author[BINP,NSU]{A.E.Bondar},
\author[BINP]{A.V.Bragin},
\author[BINP,NSU]{S.I.Eidelman},
\author[BINP,NSU]{D.A.Epifanov},
\author[BINP,NSU]{G.V.Fedotovich},
\author[BINP]{N.I.Gabyshev},
\author[BINP]{A.A.Grebenuk},
\author[BINP,NSTU]{D.N.Grigoriev},
\author[BINP,NSU]{F.V.Ignatov},
\author[BINP]{S.V.Karpov},
\author[BINP,NSU]{V.F.Kazanin},
\author[BINP,NSU]{B.I.Khazin},
\author[BINP,NSU]{I.A.Koop},
\author[BINP]{P.P.Krokovny\thanksref{where}},
\author[BINP,NSU]{A.S.Kuzmin},
\author[BINP,NSU]{I.B.Logashenko},
\author[BINP]{P.A.Lukin\thanksref{someone}},
\author[BINP]{A.P.Lysenko},
\author[BINP]{K.Yu.Mikhailov},
\author[BINP]{V.S.Okhapkin},
\author[BINP,NSU]{E.A.Perevedentsev},
\author[BINP]{A.S.Popov},
\author[BINP]{S.I.Redin},
\author[BINP]{A.A.Ruban},
\author[BINP]{N.M.Ryskulov},
\author[BINP]{Yu.M.Shatunov},
\author[BINP,NSU]{B.A.Shwartz},
\author[BINP]{A.L.Sibidanov},
\author[BINP]{I.G.Snopkov\thanksref{dec}},
\author[BINP,NSU]{E.P.Solodov},
\author[BINP]{Yu.V.Yudin},
\author[BINP]{A.S. Zaitsev}
\address[BINP]{Budker Institute of Nuclear Physics, Novosibirsk,
630090, Russia}
\address[NSU]{Novosibirsk State University, Novosibirsk, 630090,
Russia}
\address[NSTU]{Novosibirsk State Technical University, Novosibirsk, 630092,
Russia}
\thanks[where]{Currently in Heidelberg University, 69047 Heidelberg, Germany}
\thanks[someone]{Corresponding author:P.A.Lukin@inp.nsk.su}
\thanks[dec]{Deceased}
\begin{abstract}
The $\phi$(1020) meson leptonic width has been determined from the
combined analysis of 4 major decay modes of the resonance
($\phi\to K^+ K^-,K^0_LK^0_S,\pi^+\pi^-\pi^0,\eta\gamma$)
studied  with the CMD-2 detector  at the VEPP-2M $e^+e^-$ collider.
The following value has been obtained: 
$\Gamma(\phi\to e^+e^-) = 1.235\pm 0.006\pm 0.022$ keV. 
The $\phi(1020)$ meson
parameters in four main decay channels have been also recalculated:
$B(\phi\to K^+K^-) = 0.493\pm 0.003\pm 0.007$, $B(\phi\to K_LK_S) = 
0.336\pm 0.002\pm 0.006$,
$B(\phi\to\pi^+\pi^-\pi^0) = 0.155\pm 0.002\pm 0.005$, 
$B(\phi\to\eta\gamma) = 0.0138\pm 0.0002\pm 0.0002$.
\end{abstract}
\end{frontmatter}
\section{Introduction}


The simplest decay of any quarkonium vector
state occurs through its annihilation into a virtual photon,
which produces a  lepton or quark-antiquark pair. A leptonic width of a 
vector state offers a measure of wave function overlap at the origin
thus providing information about interactions of quarks composing
the vector meson.   For these reasons, decays to lepton pairs
are heavily studied and used to characterize the most basic
features of each vector state.

The present analysis is devoted to a measurement of the $\phi$(1020) 
leptonic width $\Gamma(\phi\to e^+e^-)$. Previously it has been
measured by various methods in a number of experiments. 
In Refs.~\cite{SND1,KLOE} the leptonic width was
determined in a direct study of $\phi\to e^+e^-$ and $\phi\to\mu^+\mu^-$ decays,
while in Ref.~\cite{CMD2} $\Gamma(\phi\to e^+e^-)$ was evaluated from the
simultaneous fit of four major decay modes of the $\phi$(1020) meson.

In this work we present a new measurement of the $\phi$(1020) leptonic width
using a combined fit of four main $\phi$(1020) decay modes
$\phi\to K^+K^-$~\cite{kpkm}, $\phi\to K_LK_S$~\cite{klks},
$\phi\to\pi^+\pi^-\pi^0$~\cite{3pi} and $\phi\to\eta\gamma$~\cite{etag},
studied with the CMD-2 detector~\cite{CMD2det} at the VEPP-2M $e^+e^-$
collider~\cite{vepp2m}. 

The cross sections of the processes 
$e^+e^-\to K^+K^-,~K_LK_S,~\pi^+\pi^-\pi^0,~\eta\gamma$, previously measured 
in the experiments~\cite{kpkm,klks,3pi,etag} are listed 
in Tables~\ref{tab:kpkm}~--~\ref{tab:etag}. 
The errors of the cross sections in the Tables are statistical only.

\begin{table}[p]
\begin{center}
\caption{Cross section of the process $e^+e^-\to\phi\to K^+K^-$ 
obtained in the analysis~\cite{kpkm}. 
 The errors of the cross section are statistical only.}
\vspace*{2mm}
\label{tab:kpkm}
\begin{tabular}{|c|c|c|c|}
\hline
$\sqrt{s}$, MeV  &  $\sigma$, nb & $\sqrt{s}$, MeV & $\sigma$, nb\\
\hline
\hline
   1011.566 $\pm$     0.255 &  83.57   $\pm$      5.89  &   1011.358 $\pm$     0.255 &  75.72   $\pm$      5.82  \\
   1016.124 $\pm$     0.026 &  549.20  $\pm$      7.47  &   1016.022 $\pm$     0.032 &  503.76  $\pm$      10.77 \\
   1017.016 $\pm$     0.024 &  853.45  $\pm$      10.33 &   1017.094 $\pm$     0.024 &  889.08  $\pm$      10.73 \\
   1017.970 $\pm$     0.020 &  1389.78 $\pm$      12.64 &   1018.020 $\pm$     0.050 &  1423.23 $\pm$      25.52 \\
   1019.204 $\pm$     0.018 &  2020.04 $\pm$      11.94 &   1018.886 $\pm$     0.020 &  1951.23 $\pm$      24.98 \\
   1020.102 $\pm$     0.018 &  1825.51 $\pm$      13.83 &   1019.684 $\pm$     0.020 &  1971.87 $\pm$      12.51 \\
   1020.974 $\pm$     0.020 &  1333.87 $\pm$      13.12 &   1020.722 $\pm$     0.026 &  1435.03 $\pm$      18.17 \\
   1021.808 $\pm$     0.026 &  917.57  $\pm$      12.46 &   1021.742 $\pm$     0.030 &  933.69  $\pm$      14.43 \\
   1022.752 $\pm$     0.046 &  626.16  $\pm$      15.12 &   1022.666 $\pm$     0.038 &  606.55  $\pm$      16.99 \\
   1028.332 $\pm$     0.255 &  143.80  $\pm$      10.40 &   1028.578 $\pm$     0.255 &  158.10  $\pm$      12.16 \\
   1034.061 $\pm$     0.255 &  70.50   $\pm$      7.54  &   &  \\
\hline
\end{tabular}
\end{center}
\end{table}

\begin{table}
\begin{center}
\caption{Cross section of the process $e^+e^-\to\phi\to K^0_LK^0_S$ 
obtained in the analysis~\cite{klks}. 
 The errors of the cross section are statistical only.}
\vspace*{2mm}
\label{tab:klks}
\begin{tabular}[c]{|c|c|c|c|}
\hline
$\sqrt{s}$, MeV  &  $\sigma$, nb & $\sqrt{s}$, MeV  &  $\sigma$, nb \\
\hline
\hline
   1010.272 $\pm$     0.030 &  42.21   $\pm$       5.17  & 1004.187 $\pm$     0.150 &  12.39   $\pm$       1.77  \\      
   1017.086 $\pm$     0.020 &  603.34  $\pm$       14.91 & 1011.602 $\pm$     0.072 &  56.62   $\pm$       6.87  \\      
   1018.136 $\pm$     0.018 &  999.68  $\pm$       34.84 & 1016.022 $\pm$     0.032 &  343.95  $\pm$       26.62 \\      
   1018.956 $\pm$     0.018 &  1278.75 $\pm$       32.27 & 1017.094 $\pm$     0.024 &  601.65  $\pm$       45.64 \\      
   1019.214 $\pm$     0.020 &  1329.35 $\pm$       38.80 & 1018.070 $\pm$     0.020 &  998.50  $\pm$       51.38 \\      
   1019.986 $\pm$     0.020 &  1325.08 $\pm$       28.63 & 1018.886 $\pm$     0.020 &  1317.09 $\pm$       23.21 \\      
   1020.128 $\pm$     0.020 &  1342.71 $\pm$       41.89 & 1019.684 $\pm$     0.045 &  1321.09 $\pm$       45.42 \\      
   1021.850 $\pm$     0.022 &  622.82  $\pm$       33.88 & 1020.722 $\pm$     0.026 &  999.30  $\pm$       49.45 \\      
   1023.972 $\pm$     0.020 &  292.26  $\pm$       14.91 & 1021.742 $\pm$     0.030 &  648.54  $\pm$       36.18 \\   
   1004.252 $\pm$     0.150 &  18.51   $\pm$       9.85  & 1022.666 $\pm$     0.038 &  428.05  $\pm$       27.35 \\   
   1010.722 $\pm$     0.112 &  52.96   $\pm$       7.53  & 1028.578 $\pm$     0.074 &  102.57  $\pm$       8.42  \\   
   1016.376 $\pm$     0.042 &  399.54  $\pm$       35.28 & 1004.640 $\pm$     0.150 &  13.58   $\pm$       4.59  \\      
   1017.156 $\pm$     0.026 &  600.22  $\pm$       45.78 & 1011.566 $\pm$     0.058 &  52.97   $\pm$       3.48  \\      
   1018.100 $\pm$     0.026 &  930.66  $\pm$       51.35 & 1016.124 $\pm$     0.026 &  350.79  $\pm$       28.31 \\      
   1019.040 $\pm$     0.022 &  1329.00 $\pm$       25.08 & 1017.016 $\pm$     0.024 &  560.58  $\pm$       42.85 \\      
   1020.088 $\pm$     0.020 &  1282.51 $\pm$       50.32 & 1017.970 $\pm$     0.020 &  931.61  $\pm$       49.23 \\      
   1021.020 $\pm$     0.024 &  941.38  $\pm$       46.99 & 1019.204 $\pm$     0.018 &  1354.30 $\pm$       25.21 \\      
   1021.886 $\pm$     0.046 &  620.70  $\pm$       40.29 & 1020.102 $\pm$     0.018 &  1251.84 $\pm$       49.67 \\      
   1027.820 $\pm$     0.088 &  126.74  $\pm$       10.35 & 1020.974 $\pm$     0.020 &  891.48  $\pm$       45.54 \\      
   1033.632 $\pm$     0.150 &  66.33   $\pm$       8.57  & 1021.808 $\pm$     0.026 &  606.96  $\pm$       37.01 \\      
   1039.476 $\pm$     0.150 &  37.92   $\pm$       6.23  & 1022.752 $\pm$     0.046 &  419.31  $\pm$       30.91 \\      
                            &                            & 1028.332 $\pm$     0.094 &  102.38  $\pm$       9.75  \\      
                            &                            & 1034.061 $\pm$     0.150 &  54.04   $\pm$       7.78  \\
\hline
\end{tabular}
\end{center}
\end{table}

%

\begin{table}
\caption{Cross section of the process
  $e^+e^-\to\phi\to\pi^+\pi^-\pi^0$ 
obtained in the analysis~\cite{3pi}. 
 The errors of the cross section are statistical only.}
\label{tab:3pi}
\vspace*{2mm}
\begin{center}
\begin{tabular}[c]{|c|c|c|c|}
\hline
$\sqrt{s}$, MeV  &  $\sigma$, nb & $\sqrt{s}$, MeV  &  $\sigma$, nb \\
\hline
\hline
    984.020 $\pm$      0.600  & 19.01  $\pm$      2.94    &
   1004.000 $\pm$      0.600  & 26.58  $\pm$      4.26    \\
   1010.598 $\pm$      0.120  & 61.17  $\pm$      4.92    &
   1016.076 $\pm$      0.024  & 234.66 $\pm$      15.27   \\
   1016.868 $\pm$      0.022  & 365.45 $\pm$      13.70   &
   1017.812 $\pm$      0.022  & 494.74 $\pm$      22.56   \\
   1017.538 $\pm$      0.014  & 464.25 $\pm$      14.86   &
   1018.432 $\pm$      0.016  & 622.74 $\pm$      31.47   \\
   1018.558 $\pm$      0.056  & 613.67 $\pm$      82.61   &
   1018.578 $\pm$      0.008  & 612.83 $\pm$      18.43   \\
   1018.690 $\pm$      0.020  & 592.69 $\pm$      32.47   &
   1018.694 $\pm$      0.032  & 613.72 $\pm$      47.78   \\
   1019.256 $\pm$      0.034  & 611.44 $\pm$      25.23   &
   1019.584 $\pm$      0.012  & 599.69 $\pm$      23.83   \\
   1019.776 $\pm$      0.016  & 557.75 $\pm$      17.36   &
   1020.632 $\pm$      0.014  & 382.04 $\pm$      12.58   \\
   1021.520 $\pm$      0.020  & 232.47 $\pm$      12.83   &
   1022.398 $\pm$      0.016  & 158.84 $\pm$      9.50    \\
   1027.460 $\pm$      0.064  & 18.98  $\pm$      1.65    &
   1033.442 $\pm$      0.200  & 3.66   $\pm$      0.62    \\
   1039.564 $\pm$      0.086  & .81    $\pm$      0.20    &
   1059.606 $\pm$      0.144  & .20    $\pm$      0.20    \\
   1004.000 $\pm$      0.600  & 35.90  $\pm$      3.79    &
   1010.434 $\pm$      0.120  & 66.65  $\pm$      4.93    \\
   1015.784 $\pm$      0.024  & 230.87 $\pm$      12.91   &
   1016.724 $\pm$      0.014  & 339.65 $\pm$      12.26   \\
   1017.654 $\pm$      0.008  & 466.54 $\pm$      11.50   &
   1018.828 $\pm$      0.010  & 626.24 $\pm$      12.33   \\
   1019.858 $\pm$      0.004  & 528.37 $\pm$      10.64   &
   1020.732 $\pm$      0.140  & 355.90 $\pm$      24.00   \\
   1021.710 $\pm$      0.100  & 190.74 $\pm$      13.57   &
   1023.258 $\pm$      0.022  & 104.72 $\pm$      6.69    \\
   1028.122 $\pm$      0.036  & 15.00  $\pm$      1.42    &
   1033.920 $\pm$      0.056  & 3.26   $\pm$      0.57    \\
   1039.750 $\pm$      0.126  & 0.88   $\pm$      0.21    &
   1050.092 $\pm$      0.118  & 0.10   $\pm$      0.10    \\
    984.000 $\pm$      0.600  & 16.43  $\pm$      2.22    &
   1004.000 $\pm$      0.600  & 34.64  $\pm$      3.86    \\
   1010.040 $\pm$      0.600  & 59.92  $\pm$      7.20    &
   1015.512 $\pm$      0.120  & 205.71 $\pm$      14.65   \\
   1016.812 $\pm$      0.100  & 358.05 $\pm$      17.39   &
   1017.042 $\pm$      0.080  & 352.44 $\pm$      24.37   \\
   1017.756 $\pm$      0.008  & 496.77 $\pm$      11.31   &
   1018.830 $\pm$      0.010  & 636.74 $\pm$      13.67   \\
   1019.548 $\pm$      0.012  & 597.86 $\pm$      13.08   &
   1020.070 $\pm$      0.080  & 512.23 $\pm$      62.52   \\
   1020.488 $\pm$      0.010  & 396.19 $\pm$      10.98   &
   1021.414 $\pm$      0.014  & 212.55 $\pm$      9.89    \\
   1022.516 $\pm$      0.034  & 118.91 $\pm$      7.70    &
   1027.470 $\pm$      0.040  & 18.65  $\pm$      1.47    \\
   1033.382 $\pm$      0.052  & 3.82   $\pm$      0.64    &
   1039.416 $\pm$      0.086  & 0.52   $\pm$      0.17    \\
   1049.234 $\pm$      0.122  & 0.14   $\pm$      0.14    &
   1059.006 $\pm$      0.160  & 0.48   $\pm$      0.47    \\
\hline
\end{tabular}
\end{center}
\end{table}


\begin{table}
\vspace*{2mm}
\begin{center}
\caption{Cross section of the process 
$e^+e^-\to\phi\to\eta\gamma$ with subsequent decay $\eta\to\gamma\gamma$
obtained in the analysis~\cite{etag}. 
The errors of the cross section are statistical only.}
\label{tab:etag}
\begin{tabular}{|c|c|c|c|}
\hline
$\sqrt{s}$, MeV  & $\sigma$, nb & $\sqrt{s}$, MeV  & $\sigma$, nb \\
\hline
\hline
   1003.91   &    1.70  $\pm$     0.38 &   
   1010.53   &    3.76  $\pm$     0.41 \\   
   1015.77   &    17.27 $\pm$     0.76 &   
   1016.77   &    25.39 $\pm$     0.69 \\   
   1016.91   &    27.68 $\pm$     1.24 &   
   1017.61   &    34.97 $\pm$     0.81 \\   
   1017.77   &    37.58 $\pm$     1.12 &   
   1018.58   &    54.11 $\pm$     1.47 \\   
   1018.83   &    53.66 $\pm$     0.76 &   
   1019.50   &    54.60 $\pm$     1.09 \\   
   1019.84   &    54.70 $\pm$     1.85 &   
   1020.62   &    39.41 $\pm$     0.76 \\   
   1021.54   &    24.94 $\pm$     0.91 &   
   1022.79   &    13.66 $\pm$     0.66 \\   
   1027.67   &    2.74  $\pm$     0.38 &   
   1033.67   &    0.96  $\pm$     0.36 \\   
   1039.59   &    0.46  $\pm$     0.33 &   
   1049.80   &    0.13  $\pm$     0.33 \\   
\hline
\end{tabular}
\end{center}
\end{table}

\section{Analysis}
To determine the leptonic width of the $\phi$(1020) meson, we perform a
simultaneous fit of the four $\phi$(1020) major decay modes with a leptonic 
width as a fit parameter. To fit the  experimental cross 
sections in different channels, we use the same
functions and fit parameters as in the 
corresponding dedicated studies:
\begin{eqnarray}
\sigma_{K^+K^-}(s) & = & \frac{1}{s^{5/2}}\cdot\frac{q^3(s)}{q^3(m^2_{\phi})}\cdot
\left|-\frac{m^3_{\phi}\sqrt{12\pi\cdot\Gamma_{\phi}\Gamma(\phi\to e^+e^-)B(\phi\to K^+K^-)/m_{\phi}}}
{D_{\phi}(s)} -\right.\nonumber\\
 &-& \frac{\sqrt{\Gamma_{\phi}\Gamma_{\omega}m^2_{\phi}m^3_{\omega}6\pi
 B(\omega\to e^+e^-)B(\phi\to K^+K^-)}}{D_{\omega}(s)} - \nonumber\\
 &-&\left. \frac{\sqrt{\Gamma_{\phi}\Gamma_{\rho}m^2_{\phi}m^3_{\rho}6\pi
  B(\rho\to e^+e^-)B(\phi\to K^+K^-)}}{D_{\rho}(s)}\right|^2\frac{Z(s)}
{Z(m^2_{\phi})}, \\
\sigma_{K^0_LK^0_S}(s) & = & \frac{1}{s^{5/2}}\cdot\frac{q^3(s)}{q^3(m^2_{\phi})}\cdot
\left|-\frac{m^3_{\phi}\sqrt{12\pi\cdot\Gamma_{\phi}\Gamma(\phi\to e^+e^-)B(\phi\to K^0_LK^0_S)/m_{\phi}}}
{D_{\phi}(s)} -\right.\nonumber\\
 &-& \frac{\sqrt{\Gamma_{\phi}\Gamma_{\omega}m^2_{\phi}m^3_{\omega}6\pi
 B(\omega\to e^+e^-)B(\phi\to K^0_LK^0_S)}}{D_{\omega}(s)} - \nonumber\\
 &+&\left. \frac{\sqrt{\Gamma_{\phi}\Gamma_{\rho}m^2_{\phi}m^3_{\rho}6\pi
  B(\rho\to e^+e^-)B(\phi\to K^0_LK^0_S)}}{D_{\rho}(s)}\right|^2, \\
\sigma_{\pi^+\pi^-\pi^0}(s) & = & \frac{1}{s^{3/2}}\cdot\frac{W(s)}
{W(m^2_{\phi})}\cdot
\left|\frac{\sqrt{m^3_{\phi}12\pi\cdot\Gamma_{\phi}
\Gamma(\phi\to e^+e^-)B(\phi\to\pi^+\pi^-\pi^0)}}
{D_{\phi}(s)}e^{\imath\psi_{\phi}} +\right.\nonumber\\
 &+&\left. 
\frac{W(m^2_{\phi})}{W(m^2_{\omega})}\frac{\sqrt{m^3_{\omega}12\pi\cdot\Gamma_{\omega}
\Gamma(\omega\to e^+e^-)B(\omega\to\pi^+\pi^-\pi^0)}}
{D_{\omega}(s)}\right|^2\nonumber\\
\sigma_{\eta\gamma}(s) & = & \frac{F_{\eta\gamma}(s)}{s^{3/2}}\cdot
\left|-\frac{\sqrt{m^3_{\phi}12\pi\cdot\Gamma_{\phi}
\Gamma(\phi\to e^+e^-)B(\phi\to\eta\gamma)/F_{\eta\gamma}(m^2_{\phi})}}
{D_{\phi}(s)} +\right.\nonumber\\
 &+& \frac{\sqrt{\Gamma_{\omega}^2m^3_{\omega}12\pi
 B(\omega\to e^+e^-)B(\omega\to\eta\gamma)/F_{\eta\gamma}(m^2_{\omega})}}{D_{\omega}(s)} + \nonumber\\
 &+&\left. \frac{\sqrt{\Gamma^2_{\rho}m^3_{\rho}12\pi
  B(\rho\to e^+e^-)B(\rho\to\eta\gamma)/F_{\eta\gamma}(m^2_{\rho})}}{D_{\rho}(s)}\right|^2,
\end{eqnarray} 
where s is the center-of-mass (c.m.) energy squared, 
$q = \sqrt{s/4-m^2_K}$~---~momentum of charged (neutral) kaon,
$m_V$, $\Gamma_V$ are the mass and total width of the vector meson V,
respectively, 
$D_V = m^2_V-s-\imath\sqrt{s}\Gamma_V(s)$ is the propagator of 
the vector meson V and energy dependence 
of the meson V total width is chosen according 
to~\cite{VDM}, $\Gamma(\phi\to e^+e^-)$ is the $\phi$(1020) meson 
leptonic width,  $B(V\to e^+e^-)$ is the branching ratio of the decay
$V \to e^+e^-$, $B(V\to X)$~---~branching ratio of the 
vector meson decay into a final state X. 
Here $W(s)$ is the function~\cite{VDM} describing the phase space of 
the $\pi^+\pi^-\pi^0$ final state, 
$F_{P\gamma}(s) = (\sqrt{s}\cdot(1-m^2_P/s)/2)^3$~--~phase space
factor  for the vector meson V decay into
a pseudoscalar meson P and photon, $\psi_{\phi}$~--~the phase of the
$\phi$-$\omega$ interference in the 
$\phi\to\pi^+\pi^-\pi^0$ decay channel.
 
The function Z(s) given by the relation
$$
Z(s)  =  1+\alpha\cdot\pi\cdot\frac{1+v^2}{2\cdot v},
$$
$$
v     =  \sqrt{1-\frac{4\cdot m^2_{K^{\pm}}}{s}}
$$
describes the Coulombic interaction of charged kaons in
the final state~\cite{Colomb}.
It should be mentioned  that $\rho$ and $\omega$ mesons 
are below the $K\bar K$ production threshold and
their contributions to $e^+e^-\to K\bar K$ have been calculated 
according to the SU(2) and SU(3) model 
predictions~\cite{VDM}.
The branching 
fractions of different channels, $\phi$(1020)-meson leptonic 
width, the resonance mass and total width as well as the phase of the 
$\phi$-$\omega$ interference in the 
$\phi\to\pi^+\pi^-\pi^0$ decay channel are parameters of the fit.  
To determine branching fractions of the four major $\phi$(1020)-meson decays,
we use a constraint:
\begin{eqnarray}
\sum_{X=K^+K^-,K_LK_S,3\pi,\eta\gamma}B(\phi\to X) & = & 1.0 - \sum_{X\ne 
K^+K^-,K_LK_S,3\pi,\eta\gamma}B(\phi\to X) =\nonumber\\
 & = & 0.99741\pm 0.00007\nonumber 
\end{eqnarray}

To estimate systematic errors of the parameters, one should take into 
account correlations between systematic errors of the  
$K^+K^-$ and $K^0_LK^0_S$ as well as between $\pi^+\pi^-\pi^0$ and
$\eta\gamma$ decay channels because of common contributions
(like, e.g., from luminosity measurement and radiative corrections). 
In Tables~\ref{kaons_syst},~\ref{rest_syst} contributions to 
a systematic error of each channel are presented. 
The correlated 
systematic errors for the $K^+K^-$ and $K_LK_S$ as well as for
$\pi^+\pi^-\pi^0$ and $\eta\gamma$ cross sections due to 
luminosity measurement and radiative corrections are equal to 1.1\% 
and 2.2\%, respectively. A difference in systematic 
errors due to luminosity measurement is caused by different detector 
conditions during  data taking periods.

\begin{table}
\begin{center}
\caption{Contributions to the systematic errors of 
$\phi\to K^+K^-$ and $\phi\to K_LK_S$ cross sections.
Common contributions of both errors are denoted with $\star$. }
\label{kaons_syst}
\begin{tabular}{|c|c|c|}
\hline
Source & $K^+K^-$  & $K_LK_S$ \\
\hline
\hline
Luminosity measurement $^{\star}$           & 1.0     & 1.0 \\
Radiative corrections$^{\star}$ & 0.5     & 0.5 \\
Selection criteria             & 1.6     & 1.2  \\
Trigger efficiency             & 1.0     & 0.5 \\
Background shape               & 0.4     & 0.3 \\
Uncertainty in energy spread   & 0.2     & 0.2 \\
\hline
\hline
Total                          & 2.2     & 1.7 \\
\hline
\end{tabular}
\end{center}
\end{table}

\begin{table}
\begin{center}
\caption{Contributions to the systematic error 
of $\phi\to \pi^+\pi^-\pi^0$ and $\phi\to\eta\gamma $cross 
sections. Common contributions of both errors are denoted with $\star$.}
\label{rest_syst}
\begin{tabular}{|c|c|c|}
\hline
Source & $\pi^+\pi^-\pi^0$  & $\eta\gamma$   \\
\hline
\hline
Luminosity measurement$^{\star}$                 & 2   & 2 \\
Radiative corrections$^{\star}$       & 1   & 1 \\
Selection criteria                   & -   & 4 \\
Trigger efficiency                   & 1   & 2 \\
Simulation statistic                 & 0.4 & - \\
Background subtraction               & 0.3 & 3 \\
$\pi^0$ reconstruction                &     &   \\
efficiency                           & 0.4 & - \\
Model uncertainty                    & -   & 0.1\\                     
\hline
\hline
Total                                & 2.5 & 5.6\\  
\hline
\end{tabular}
\end{center}
\end{table}

To determine  the leptonic width and 
branching fractions taking into account systematic errors, we use the
Maximum Likelihood method with the 
following likelihood function:
$$
\mathcal{L} = -\left(\sum_{\imath}
 \frac{(f^{data}_{\imath}\cdot(1+\delta_{j})\cdot(1+\Delta_k)-f^{theor}_{\imath})^2}{2\sigma^2_{\imath}}
+\frac{\delta^2_{j}}{2\delta^2_{est,j}}
+\frac{\Delta^2_k}{2\Delta^2_{est,k}}\right),
$$
where $f^{data}_{\imath}$ is the experimental value of the cross
section for the process $\imath$
($\imath = K^+K^-$, $K_LK_S$, $\pi^+\pi^-\pi^0$, $\eta\gamma$), 
$f^{theor}_{\imath}$ --the value of the 
theoretical cross section for the process $\imath$, subscript $j$ counts
an ``individual'' part of a systematic error in the cross section 
of the process $\imath$ and $\Delta_1$ denotes a common 
part of the systematic error in measurements of the kaon cross
sections,  while $\Delta_2$ means a common part of 
systematic errors for the $\pi^+\pi^-\pi^0$ and $\eta\gamma$ studies. 
The following values have been 
obtained from the maximization procedure: 
\begin{eqnarray}
B(\phi\to K^+K^-) & = & 0.493 \pm 0.008\nonumber \\
B(\phi\to K^0_LK^0_S) & = & 0.336\pm 0.006\nonumber \\
B(\phi\to \pi^+\pi^-\pi^0) & = & 0.155\pm 0.005\nonumber \\
B(\phi\to\eta\gamma) & = & 0.0138\pm 0.0003\nonumber \\
m_{\phi} & = & 1019.437\pm 0.010\mbox{ MeV/c}^2 \nonumber \\
\Gamma_{\phi} & = & 4.220\pm 0.025\mbox{ MeV}\nonumber \\
\Gamma_{ee} & = & 1.206\pm 0.022 \mbox{ keV}\nonumber\\
\chi^2/n.d.f. & = & 116.50/130\nonumber,
\end{eqnarray}
where errors of the parameters are experimental  (i.e., include 
statistical and systematic uncertainties). 

To determine the statistical errors of the parameters separately, 
the same fit has been performed with $\Delta_k$ and 
$\delta_j$ fixed at zero. The following values have been obtained:
\begin{eqnarray}
B(\phi\to K^+K^-) & = & 0.494 \pm 0.003\nonumber \\
B(\phi\to K^0_LK^0_S) & = & 0.335\pm 0.002\nonumber \\
B(\phi\to \pi^+\pi^-\pi^0) & = & 0.154\pm 0.002\nonumber \\
B(\phi\to\eta\gamma) & = & 0.0140\pm 0.0002\nonumber \\
m_{\phi} & = & 1019.437\pm 0.007\mbox{ MeV/c}^2 \nonumber \\
\Gamma_{\phi} & = & 4.220\pm 0.019\mbox{ MeV}\nonumber \\
\Gamma_{ee} & = & 1.219\pm 0.006 \mbox{ keV}\nonumber.
\end{eqnarray}

As one can see, the central values of the parameters from the last fit  
are slightly shifted with respect to 
the results of the previous fit. Using Monte-Carlo simulation it was 
checked that a variation of the shape of the likelihood function 
shifts the "true" value of the leptonic width by 
$-(0.0051\pm 0.0001)$ keV, while taking into account correlations 
between the systematic errors
 leads to changing the $\Gamma_{ee}$ input value by $-(0.0287\pm 0.0002)$ keV.
So, the obtained value of the 
$\Gamma_{ee} = 1.206\pm 0.022$ keV should be corrected by this shift.
Thus our final result for the $\phi$(1020) leptonic width is:
$$
\Gamma_{ee} = 1.235\pm 0.022\mbox{ keV.}
$$

The assumption of SU(2) and SU(3) symmetry~\cite{VDM} used to calculate
the   $\rho,\omega\to K\bar K$ contributions is valid within 
$\sim$20\% accuracy.
To estimate a systematic error due to the choice 
of the fitting model we performed a fit
with the $\phi$(1020) contribution only in the $e^+e^-\to K\bar K$ 
channels. The obtained differences in the values of the fitting 
parameters were less than 0.5\% and used as a
model systematic error. 

Contributions to the systematic error due to uncertainties in
parameters used as fit constants 
($m_{\rho}$, $m_{\omega}$, $\Gamma_{\rho}$, $\Gamma_{\omega}$, ...)
are at the level of 10$^{-5}$.

\begin{figure}
\includegraphics[width=0.9\textwidth]{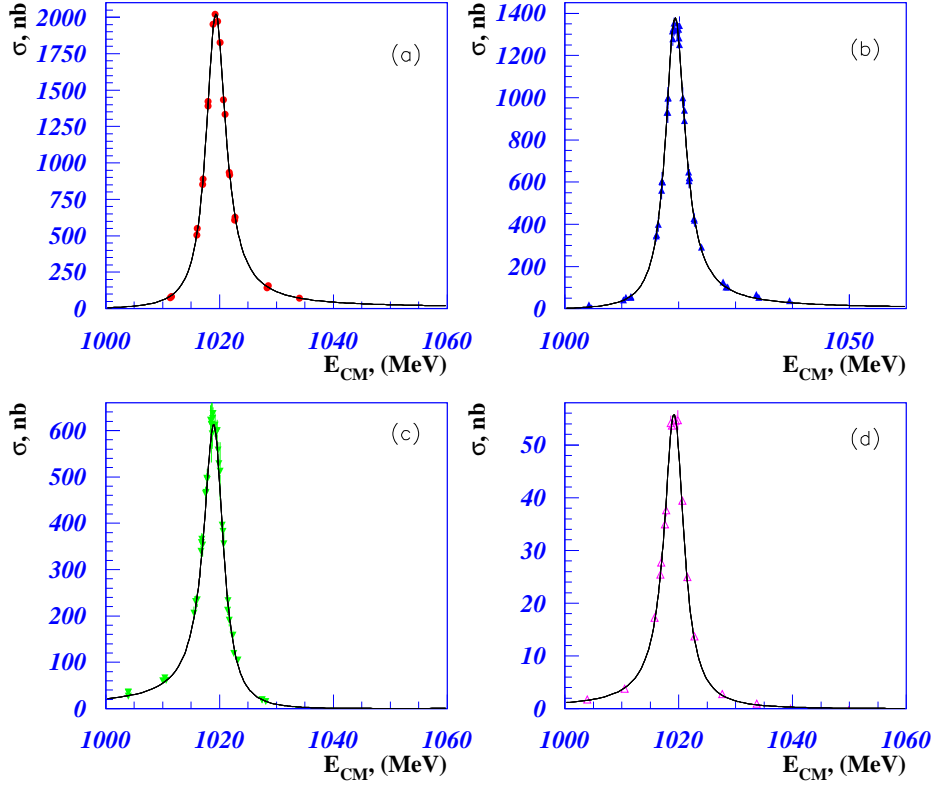}
\caption{Energy dependence of the cross sections 
for the processes $\phi\to K^+K^-$(a), $\phi\to K^0_LK^0_S$(b), 
$\phi\to\pi^+\pi^-\pi^0$(c) and $\phi\to\eta\gamma$(d).}
\label{phi4modes}
\end{figure}

\begin{figure}
\includegraphics[width=0.9\textwidth]{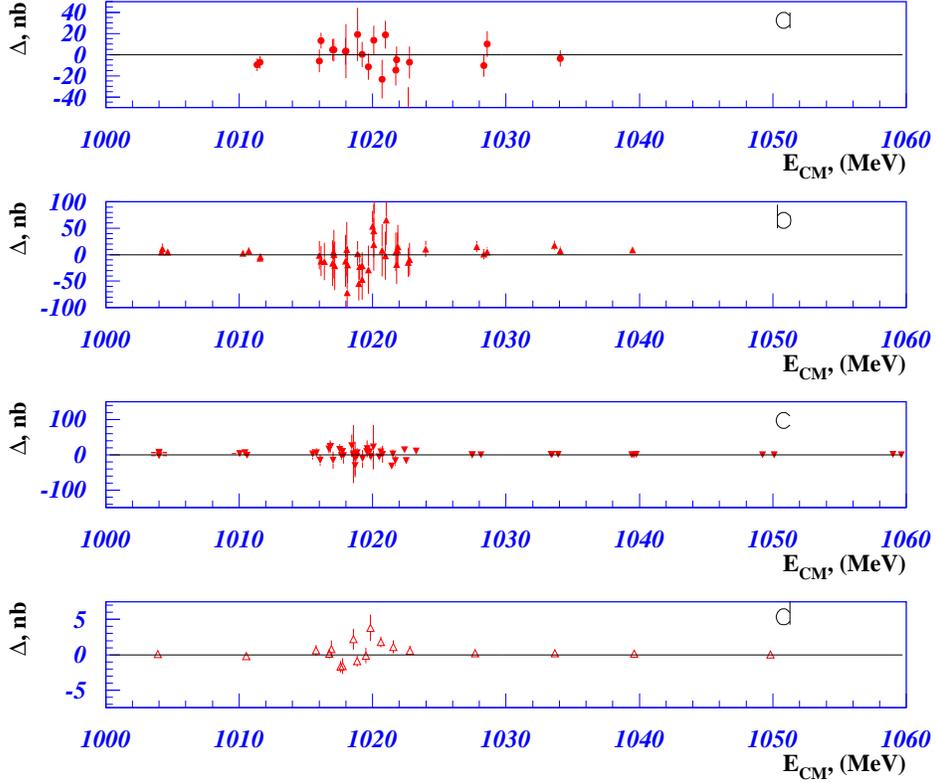}
\caption{Residuals between the measured cross section 
and theoretical one, calculated as a function of c.m. energy for
the $\phi\to K^+K^-$ (a), $\phi\to K^0_LK^0_S$ (b), 
$\phi\to\pi^+\pi^-\pi^0$ (c), and 
$\phi\to\eta\gamma$ (d) decays.}
\label{deltas}
\end{figure}

In plots of Fig.~\ref{phi4modes}(a--d) one can see the c.m. energy 
dependence of the cross sections for the processes under study 
along with the corresponding fitting curves. 
In Fig.~\ref{deltas}(a--d) the differences between the cross sections 
and the values of the fitting curves for all the processes are presented.

So, our final results are:
\begin{eqnarray}
B(\phi\to K^+K^-) & = & 0.493 \pm 0.003\pm 0.007\nonumber \\
B(\phi\to K^0_LK^0_S) & = & 0.336\pm 0.002\pm 0.006\nonumber \\
B(\phi\to \pi^+\pi^-\pi^0) & = & 0.155\pm 0.002\pm 0.005\nonumber \\
B(\phi\to\eta\gamma) & = & 0.0138\pm 0.0002\pm 0.0002\nonumber \\
m_{\phi} & = & 1019.437\pm 0.007\pm 0.007\mbox{ MeV/c}^2 \nonumber \\
\Gamma_{\phi} & = & 4.220\pm 0.019\pm 0.016\mbox{ MeV}\nonumber \\
\Gamma_{ee} & = & 1.235\pm 0.006\pm 0.022\mbox{ keV}\nonumber,
\end{eqnarray}
where the first error  is statistical and the second 
is systematic.

\begin{figure}
\includegraphics[width=0.9\textwidth]{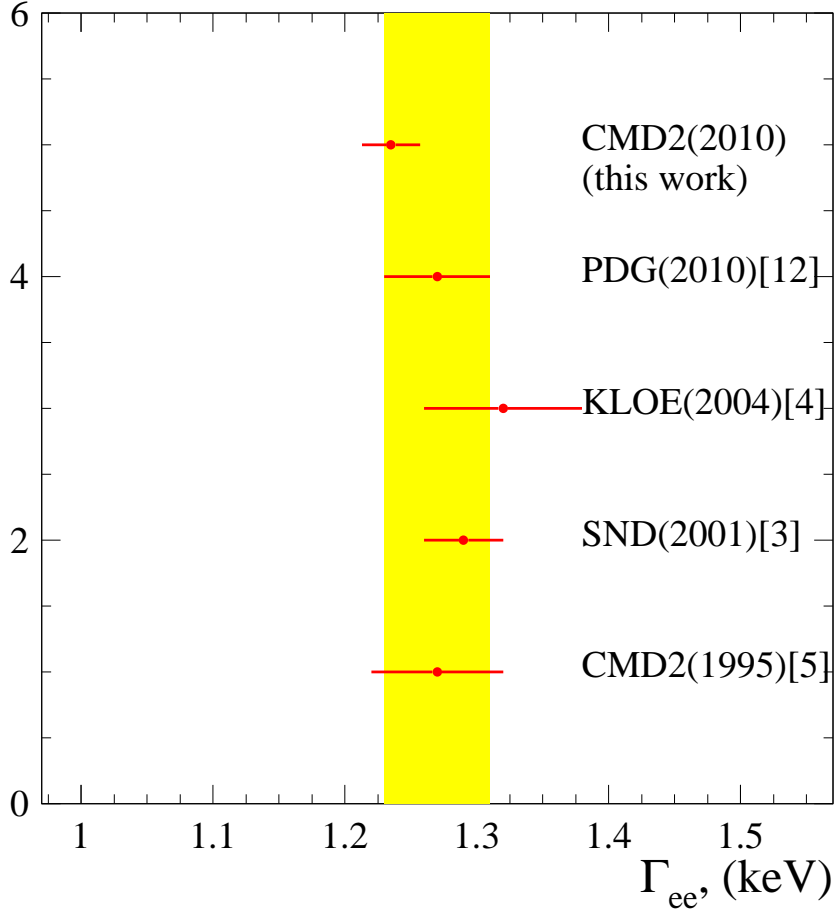}
\caption{The results of $\Gamma_{ee}$ measurements by
CMD-2~\cite{CMD2}, SND~\cite{SND1} and  KLOE~\cite{KLOE} 
as well as the result of the present analysis. The 
shaded region represents the PDG evaluation 
with its error~\cite{PDG}. }
\label{gee_exps}
\end{figure}

In Fig.~\ref{gee_exps} the results of different  
measurements of $\Gamma_{ee}$ are shown
along with the result of the present analysis. The shaded region 
corresponds to the leptonic width value from~\cite{PDG}
with its accuracy. As can be seen, the result of our analysis 
is in good agreement with results of other 
measurements and has better precision.
\vspace*{0.5cm}
\section{Conclusion}
From combined analysis of four major $\phi$(1020) meson decay modes 
the leptonic width $\Gamma(\phi\to e^+e^-)$ 
has been measured:
$$
\Gamma_{ee} = 1.235\pm 0.006\pm 0.022\mbox{ keV.}
$$
The measurement is the most precise one obtained by now. 
The precision of four $\phi$(1020) major decay
modes has been improved. The following values of the branching
fractions have been obtained:

\begin{eqnarray}
B(\phi\to K^+K^-) & = & 0.493 \pm 0.003\pm 0.007\nonumber \\
B(\phi\to K^0_LK^0_S) & = & 0.336\pm 0.002\pm 0.006\nonumber \\
B(\phi\to \pi^+\pi^-\pi^0) & = & 0.155\pm 0.002\pm 0.005\nonumber \\
B(\phi\to\eta\gamma) & = & 0.0138\pm 0.0002\pm 0.0002\nonumber.
\end{eqnarray}


The obtained value of the $\phi$(1020) leptonic width is smaller than 
the value in the previous CMD-2 
measurement~\cite{CMD2} by about one experimental error 
reflecting a decrease of the 
total width of the $\phi$(1020) meson. 

The value of the $\phi$(1020) meson leptonic width obtained here is
correlated to the values of the four major 
$\phi$(1020) branching fractions and therefore should not be 
included in the constrained fit performed by PDG.

All parameters ($\Gamma_{ee}$ and $B(\phi\to X)$) are in  
good agreement with results of other experiments.


\section{Acknowledgments}
\hspace*{\parindent} The authors are grateful to the staff of VEPP-2M for
excellent collider performance, to all engineers and technicians who
participated in the design, commissioning and operation of CMD-2.

\par This work is supported in part by FEDERAL TARGET PROGRAM "Scientific
  and scientific-pedagogical personnel of innovative Russia in 2009-2013"
  and by the grants \mbox{INTAS YSF 06-100014-9464}, 
\mbox{INTAS 1000008-8328},\\ \mbox{DFG GZ 436 RUS 113/769/0-3}, 
\mbox{RFBR 10-02-00253}, \mbox{RFBR 10-02-00695}.

\end{document}